\documentstyle[tighten,preprint,eqsecnum,aps,prd]{revtex}

\def\no{\nonumber}
\def\be{\begin{equation}}
\def\ee{\end{equation}}

\def\e{\epsilon}
\def\m{\mu}
\def\n{\nu}
\def\t{\theta}

\begin{document}
\draft

\preprint{\vbox{\baselineskip=12pt
\rightline{IUCAA-17/99}
\rightline{}
\rightline{hep-th/0001119}
}}
\title{The electrogravity transformation and global monopoles in 
scalar-tensor gravity}
\author{Sukanta Bose \footnote{Electronic address:
{\em sbose@iucaa.ernet.in}} and Naresh Dadhich\footnote{Electronic address: 
{\em nkd@iucaa.ernet.in}}}
\address{Inter-University Centre for Astronomy and Astrophysics, Post Bag 4,
Ganeshkhind,\\ Pune 411007, India}
\date{April 1999}

\maketitle
\begin{abstract}
 
The electrogravity transformation is defined by an interchange of the 
``active'' and ``passive'' electric parts of the Riemann tensor. Such a 
transformation has been used to find new solutions that are ``dual'' to the 
Kerr family of black hole spacetimes in general relativity. In such a case, 
the dual solution is a similar black hole spacetime endowed with a global 
monopole charge. Here, we extend this formalism to obtain solutions dual to
the static, spherically symmetric solutions of two different scalar-tensor 
gravity theories. In particular, we first study the duals of the charged black 
hole solutions of a four-dimensional low-energy effective action of heterotic 
string theory. Next, we study dual of the Xanthopoulos-Zannias solution in 
Brans-Dicke theory, which contains a naked singularity. We show that, 
analogous to general relativity, in these scalar-tensor gravity theories the 
dual solutions are similar to the original spacetimes, but with a 
global monopole charge.

\end{abstract}

\narrowtext

\vfil
\pagebreak

\section{Introduction}
\label{sec:intro}

Ever since the work of Barriola and Vilenkin (BV) \cite{BV}, it has been of 
interest to explore the existence of spacetimes with a global monopole charge 
in alternative theories of gravity. The motivation for considering gravity 
theories other than general relativity (GR), which was the domain of the BV 
paper, is that topological defects such as global monopoles are believed to 
have formed as an inevitable consequence of phase transitions that occurred 
in the early universe. There is also reason to believe that (GR)
may not necessarily be the only viable theory of gravity. Therefore, it is 
interesting to ask if the global monopole spacetimes in viable alternative 
gravity theories are sufficiently different from the BV solution. On the other
hand, it is also important to study if there are any common underlying 
features shared by these spacetimes arising as solutions in different gravity 
theories. Indeed this constitutes one of the main motivations for this paper. 
Here, we show that in some scalar-tensor theories of gravity, spacetimes with
a global monopole charge, which by definition are static and spherically 
symmetric (SSS), arise from other SSS solutions by an electrogravity  
transformation \cite{ND1} defined below. 

It is well known that in general relativity, in analogy with Maxwell's theory 
of electromagnetism, one can extract the ``electric'' and ``magnetic'' parts 
of the gravitational field from the Riemann curvature tensor \cite{CL,LB,BS}. 
These parts are defined relative to a timelike unit vector field and are 
represented by second rank tensors in three-dimensional space transverse to 
that field. There are two types of electric parts, ``active'' and ``passive''. 
The former is obtained by projecting two of the indices of the Riemann tensor 
along the timelike unit vector, while the latter is obtained by a similar 
operation on the double-dual of the Riemann tensor (see section \ref{sec:eg} 
below). Between them, the active and passive parts comprise of 6 independent 
components each since they are given by symmetric tensors. An analogous 
projection of the left dual of the Riemann tensor yields a 
traceless second rank tensor in three space dimensions, which is called the 
magnetic part. It has 8 independent components. Therefore, the two electric 
parts and the magnetic part completely determine the Riemann tensor. 

In general relativity (GR), a field transformation analogous to the 
electromagnetic-duality transformation in Maxwell's theory can be defined that 
keeps the Einstein-Hilbert action invariant. Such a transformation 
simultaneously maps the active electric part and the magnetic part into the 
magnetic part and minus the passive electric part, respectively. In fact, such 
a set of transformation equations implies the Einstein vacuum field-equations, 
with a vanishing cosmological constant \cite{ND1,ND2}. The 
electromagnetic-duality transformation in Maxwell's theory does not exhibit 
such a property since it involves only the fields, which are defined in terms 
of only the first derivative of the gauge potential. By contrast, the electric 
and magnetic parts of the gravitational field are obtained from projections of 
the Riemann tensor and its duals, and hence, contain second-order derivatives 
of the metric. These parts, therefore, can be appropriately combined to lay 
down the equations governing the dynamics of the gravitational field.

Another kind of transformation, termed as the electrogravity transformation
(EGT), can be defined by interchanging the active  part of the gravitational 
field with the passive one \cite{ND3}. Such a transformation is a symmetry of 
the Einstein-Hilbert action: It maps this action into itself, provided one 
simultaneously transforms the gravitational constant as, $G\rightarrow -G$. 
Therefore, the vacuum field equations are left invariant under EGT 
\cite{ND3}. In this sense it is like a duality transformation. The need to 
implement the change in sign of $G$ in the ``dual'' solutions can be 
understood as follows. The source term for the active electric part stems 
from the matter stress tensor and can be argued to represent matter energy, 
while the passive part is associated with the gravitational-field energy. For 
an attractive field, these two kinds of energy must bear opposite signs. Thus, 
$G$ has to change sign under the interchange of active and passive electric 
parts. 

The field equations coupled to matter, however, are not invariant under EGT.
By a common abuse of language, however, we will refer to the field equations
transformed under the EGT as the ``dual'' equations, and their solutions as
the ``dual'' solutions.

Importantly, EGT can be effectively used to obtain new solutions from known 
ones. In this context, note that EGT transforms the Ricci tensor into the 
Einstein tensor, and vice versa. This is because contraction over a pair of 
Riemann tensor indices yields the Ricci tensor while a similar contraction on 
its double dual yields the Einstein tensor.  For all vacuum solutions, it is 
possible to introduce matter terms suitably in the field equations such that 
the modified field equations still admit the original vacuum solutions. 
However, since the field equations coupled to matter are not EGT invariant, 
the solutions to the dual equations will be, in 
general, different from the original vacuum solutions. This is what happens
for the static, spherically symmetric family of black hole solutions in GR. In 
fact, it has been shown that in such a case, a typical dual solution represents
a black hole endowed with a global monopole charge. Here, we extend this 
formalism to obtain solutions dual to the static, spherically symmetric 
solutions of Brans-Dicke theory as well as a 
four-dimensional (4D) low-energy effective action of heterotic string theory, 
namely, dilaton gravity coupled to a $U(1)$ gauge field. We show that 
analogous to general relativity, in these scalar-tensor gravity theories the 
dual solutions are similar to the original spacetimes, but with a 
global monopole charge, which, therefore, appears to be a rather generic 
feature of EGT. Our work also implies 
that the effect of a spontaneous symmetry breaking on the global monopole 
field is tantamount to performing an EGT on 
the active and passive electric components of the spacetime curvature.

The layout of the paper is as follows. In section \ref{sec:eg}, we define the 
electrogravity transformation and show how the vacuum Einstein field
equations remain unchanged under it. We briefly recapitulate how the solution 
dual to Schwarzschild can be obtained by implementing this transformation,
in section \ref{sec:sd}. The global monopole field configuration and the 
associated matter stress tensor are discussed in section \ref{sec:gm}. In 
section \ref{sec:dg}, we discuss the charged black hole solutions of dilaton 
gravity and find its dual by effecting the EGT. In section \ref{sec:bd}, we 
obtain the dual of the Xanthopoulos-Zannias solution in Brans-Dicke theory. A 
few thoughts on these solutions and scope for future work are presented in 
section \ref{sec:disc}. We work with the metric
signature $(-,+,+,+)$ and employ geometrized units $G=1=c$.
 
\section{Electrogravity duality}
\label{sec:eg}
 
The electric and magnetic parts of the gravitational field in general 
relativity are defined as follows. Consider a timelike unit vector field $u^a$,
with $u^a u_a = -1$.
Then the active and passive parts of the Riemann tensor relative to $u^a$ are
\be\label{el}
E_{ac} = R_{abcd} u^b u^d, \quad \tilde E_{ac} = *R*_{abcd} u^b u^d \ \ ,
\ee
respectively. Above, $*R*_{abcd}$ is the double-dual of the Riemann tensor 
given by:
\be\label{dd}
*R*_{abcd} = {1\over 4} \e_{abef}\e_{cdgh} R^{efgh} \ \ ,
\ee
where $\e_{abcd}$ is the canonical four-volume element of the spacetime. The 
magnetic part is the projection of left or right dual of the Riemann 
tensor and is given by 
\be\label{mag}
H_{ac} = -*R_{abcd} u^b u^d = H_{(ac)} - H_{[ac]} \ \ ,
\ee
where we have used the left-dual,
\be\label{ld}
*R_{abcd} = {1\over 2} \e_{abef} R^{ef}_{\>\>cd} \,.
\ee
Also, the symmetric and antisymmetric parts of $H_{ac}$ can be expressed as: 
\be\label{Has}
H_{(ac)} = -*C_{abcd} u^b u^d \quad {\rm and} \quad H_{[ac]} = -\frac{1}{2} 
\e_{abce} R^e_d u^b u^d \ \ ,
\ee
where $C_{abcd}$  is  the  Weyl tensor. Thus, the symmetric part is equal to 
the Weyl magnetic part, whereas the anti-symmetric part represents energy flux.
Note that $E_{ab}$ and ${\tilde E}_{ab}$ are both symmetric while $H_{ac}$ is 
trace-free and they are all purely spacelike, i.e., $(E_{ab}, {\tilde E}_{ab}, 
H_{ab})u^b = 0$. The Ricci tensor can then be expressed in terms of the 
electric and magnetic parts as
\be\label{Rem}
R_a{}^b = E_a{}^b + {\tilde E}_a{}^b - (E + {\tilde E}) u_a u^b - {\tilde E } 
\delta_a{}^b - (\e_{amn} H^{mn} u^b + \e^{bmn} H_{mn} u_a)
\ee
where $E = E_a{}^a$ and $\tilde E = \tilde E_a{}^a$. 

The EGT is defined by an interchange of the active and passive parts of the 
electric field, and mapping the magnetic part to minus itself:
\be \label{egd}
E_{ab} \longleftrightarrow {\tilde E}_{ab}, ~H_{ab} \longrightarrow  -H_{ab}.
\ee
To see the effect of this transformation on vacuum solutions, note that the
vacuum field equations, $R_{ab} = 0$, are in general equivalent to  
\be\label{vac}
E ~ {\rm or} ~ {\tilde E} = 0,~ H_{[ab]} = 0 = E_{ab} + {\tilde E}_{ab}
\ee
which are symmetric in $E_{ab}$ and ${\tilde E}_{ab}$. Thus the vacuum field
equations (\ref{vac}) are invariant under EGT (\ref{egd}). 

\section{Schwarzschild dual}
\label{sec:sd}

To set the notation and to aid the discussion of SSS solutions and their duals
in scalar-tensor gravity, we briefly study how one arrives in GR at the dual 
of the Schwarzschild solution \cite{ND3}. Birkoff's theorem implies that the 
Schwarzschild solution, characterized by its mass, is the unique spherically 
symmetric solution to Einstein's vacuum field equations. Any spherically 
symmetric metric can be cast in the form:
\be\label{SSS}
ds^2 = - e^{2\nu} dt^2 + e^{2\lambda} dr^2 + h^2 d \omega^2 \ \ ,
\ee
where $\nu$, $\lambda$, and $h$ are functions of time, $t$, and the radial 
coordinate $r$, and $d\omega^2$ is the line element on a unit two-sphere.
It is well known that under the above conditions we can choose a gauge where
$h=r$. This is what we do first. A natural choice for the timelike vector 
$u^a$ in this case is the timelike unit normal to $t=$constant hypersurfaces. 
Then, a subset of the conditions in Eqs. (\ref{vac}) that ensure a vacuum 
solution, namely, $H_{[ab]} = 0$ and $E_\t{}^\t + {\tilde E}_\t{}^\t = 0$ 
imply that $\nu +\lambda = 0$. A supplementary requirement of ${\tilde E} = 0$
yields $e^{-2\lambda} = (1-2M/r)$. This leads to the Schwarzschild solution. 
Here, it is important to realize that we were not required to impose the 
remaining condition in Eqs. (\ref{vac}), namely, $E_r{}^r + {\tilde E}_r{}^r 
= 0$, in order to obtain this solution. In fact, this equation 
is implied by the rest. We thus have a choice for introducing some matter
distribution in the $r$-direction without affecting the Schwarzschild 
solution. We modify the vacuum field equations (\ref{vac}) to read as
\be\label{modvac}
H_{[ab]} = 0 = {\tilde E}\quad {\rm and} \quad E_{ab} + {\tilde E}_{ab}
= k w_a w_b 
\ee
where $k$ is a scalar and $w_a$ is a spacelike unit vector along the radial 
acceleration vector $\dot u_a = u^b\nabla_b u_a$. It is clear that the above 
equations once again admit the Schwarzschild solution as the unique 
spherically symmetric solution with $k = 0$. 
    
We now perform the electrogravity transformation (\ref{egd}) on the above set 
of equations (\ref{modvac}) to obtain:  
\be\label{dmodvac}
H_{[ab]} = 0 = E \>, \quad E_{ab} + {\tilde E}_{ab} = k w_a w_b \,.
\ee
Its general solution is given by the metric (\ref{SSS}) with
\be\label{gms}
e^{2\nu} = e^{-2\lambda} = \left(1 - 8\pi\eta^2 - \frac{2M}{r}\right) \ \ ,
\ee
which is the Barriola-Vilenkin solution \cite{BV} for a Schwarzschild particle 
with global monopole charge parameter, $\sqrt {2k}$. This solution is obtained
as follows. The condition $\nu +\lambda = 0$ is implied by the equation 
$E_\t{}^\t + {\tilde E}_\t{}^\t = 0$. In addition to this, the condition 
$E=0$ yields $e^{2\nu} = (1- 8\pi\eta^2- 2M/r)$ and $k = 4\pi\eta^2 /r^2$. 
This spacetime has non-zero stresses given by
\be\label{gmstress}
8\pi T_t{}^t = 8\pi T_r{}^r = \frac{2k}{r^2}.
\ee
Just like the Schwarzschild solution, the monopole solution (\ref{gms}) is also
the unique solution of Eq. (\ref{dmodvac}). 

One may now ask if there are any common features that the dual solutions 
share. This can be easily found for spherically symmetric solutions with the 
metric (\ref{SSS}). For such solutions, the condition $E_\t{}^\t + 
{\tilde E}_\t{}^\t = 0$ from Eq. (\ref{modvac}) is tantamount to $R_t{}^t= 
R_r{}^r$ in terms of the Ricci-tensor components. It implies that $\nu+
\lambda= 0$. Together with $\tilde{E}=0$ this implies that $R_\t{}^\t=0= 
R_\phi {}^\phi$. The other components of $R_a{}^b$ are zero owing to the 
remaining condition in Eq. (\ref{modvac}), namely, $H_{[ab]} =0$. These 
conditions on the components of the Ricci tensor can be satisfied even by a 
matter stress tensor that does not necessarily represent vacuum. In 
fact, a particular example of $R_{ab}$ conforming to these requirements is
\be\label{Rgen}
R_a{}^b = k \left(w_a w^b - u_a u^b \right) \,.
\ee
The matter stress tensor associated with the above form of the Ricci tensor is 
\be\label{Tgen}
8\pi T_a{}^b = k \left(w_a w^b - u_a u^b -\delta_a{}^b\right) \ \ ,
\ee
where, in general, $k$ can be a function of $r$ and $t$.

Since the EGT interchanges $R_a{}^b$ with $G_a{}^b$, the dual spacetimes are 
solutions of Eq. (\ref{Rgen}), with $R_a{}^b$ replaced by $G_a{}^b$. Hence, 
they are solutions to a different matter distribution, given by
\be\label{Tdgen}
8\pi T_a{}^b = k \left(w_a w^b - u_a u^b \right) \,.
\ee
In this case, the Einstein field equation implies that 
\be\label{Rdgen}
R_a{}^b = k \left(w_a w^b - u_a u^b -\delta_a{}^b \right) \,.
\ee
Consequently, the equations of motion are:
\begin{mathletters}%
\label{Rtrgen}
\begin{eqnarray}
R_t{}^t =R_r{}^r &=& 0 \ \ , \label{Rttgen} \\
R_r{}^t &=& 0\ \ , \label{Rrtgen} \\
R_\t{}^\t &=& k \,. \label{Rthgen}
\end{eqnarray}
\end{mathletters}%
As before, Eq. (\ref{Rttgen}) implies $\nu+\lambda =0$. However, for a 
non-vanishing $k$, we can expect $h$ to be different from $r$ here, unlike in 
the case of the Schwarzschild solution. Since
\be\label{Rth}
R_{\t\t} = \left\{1+e^{-2 \lambda} \left[r(\lambda'-\nu') -1\right]\right\} +
e^{-2\lambda} \left[(hh'- r) (\lambda'-\nu') - (hh''+h'^2) + 1\right] \ \ ,
\ee
one immediately notices that $k \propto 1/r^2$ will always act as a 
source for the above Ricci tensor component if $h= {\rm const}\times r^2$, 
where the {\em constant} is different from 1. In fact, for SSS line-elements 
in general relativity this gives rise to a general prescription to obtain dual 
solutions by changing $h$ in the above manner \cite{DP1}. 

\section{Global monopole}
\label{sec:gm}

Global monopoles are stable topological defects. They are supposed to be 
produced when global symmetry is spontaneously broken in phase transitions in 
the early Universe \cite{VS}. A global monopole is described by an isoscalar 
triplet, $\psi^a$, with $a=1,\>2,\>3$. The associated Lagrangian density is 
\cite{BV}:
\be\label{Lgm}
L_m = {1\over 2} (\nabla \psi^a )^2 +{\lambda\over 4}(\psi^a\psi_a -\eta^2 )^2 
\ \ . \ee
Such a system has a global $O(3)$ symmetry and offers topologically 
non-trivial self-supporting solutions.
The global monopole is obtained by implementing the ansatz that $\psi^a (r)=
\eta f(r)x^a /r$, where $x_a x^a = r^2$. Here $\eta$ is a constant whose value 
defines the energy scale of symmetry breaking. 

For a given spacetime metric, the stress tensor associated with a global 
monopole can be inferred from the above Lagrangian density in a standard 
manner. Consider the SSS metric (\ref{SSS}) with $\nu$ and $\lambda$ 
independent of $t$. Then $x^a$ is interpreted as a ``Cartesian'' coordinate, 
and the field equation for $\psi^a$ reduces to the following equation for 
$f(r)$.
\be\label{fr}
e^{-2\lambda} f'' +\left[{2e^{-2\lambda}\over r} + {e^{-2\nu}\over 2}\left( 
e^{2(\nu-\lambda)} \right)^{'} \right]f' -{2f\over r^2} -\lambda \eta^2 f(f^2 
-1) =0 \,.
\ee
The stress-tensor components of the monopole are:
\begin{eqnarray}
T_t{}^t &=& {\eta^2 f'^2 e^{-2\lambda} \over 2}+{\eta^2 f^2 \over r^2}+{1
\over 4}\lambda\eta^4(f^2-1)^2 \ \ ,
\no \\
T_r{}^r &=&-{\eta^2 f'^2 e^{-2\lambda}\over 2}+{\eta^2 f^2 \over r^2}+{1
\over 4}\lambda\eta^4(f^2-1)^2 \ \ ,
\no \\
T_\t{}^\t &=& T_\phi{}^\phi = {\eta^2 f'^2 e^{-2\lambda}\over 2}+{1\over 4}
\lambda\eta^4(f^2-1)^2 \,.
\label{gmst}\end{eqnarray}
The monopole core is defined by values of $r$ for which $f(r)\approx 1$.
Outside and at large distances from the 
monopole core  the stresses would approximate to \cite{AV} 
\be\label{emt}
T_r{}^r \approx T_t{}^t \approx {\eta^2\over r^2} \>, \quad T_\t{}^\t = 
T_\phi{}^\phi\approx 0 \ \ ,
\ee
which is precisely of the form given in Eq. (\ref{gmstress}). 

The dual solution to flat spacetime can also be obtained as follows. Note that 
flat spacetime is a solution to the following equations of motion:
\be \label{eomf}
{\tilde E}_{ab} = 0 = H_{[ab]}\>, \quad E_{ab} = k w_a w_b \ \ ,
\ee
which are solved to give $\nu=\lambda=0$. As before, the condition 
${k} = 0$ 
is implied by the fact that such a solution corresponds to an isotropic 
spacetime. Its dual is the solution of the equation dual to (\ref{eomf}),
which reads as 
\be
E_{ab} = 0 = H_{[ab]}, \quad {\tilde E}_{ab} = k w_a w_b
\ee
yielding the general solution,
\be
\nu^{\prime} = \lambda^{\prime} = 0 \Longrightarrow e^{2\nu}=1, \>e^{2\lambda}
= (1-2k)^{-1} ={\rm constant} \,. \ee
The resulting spacetime is non-flat and represents a global monopole of zero  
mass. Note that such a spacetime is the same as the one described by 
Eq. (\ref{gms}) in the limit of vanishing mass $M$. This could as well be 
considered as a spacetime of constant relativistic potential. It can also
be viewed as a ``minimally" curved spacetime (see Refs. \cite{ND2,ND3}).
The EGT thus generates topological defects in vacuum solutions of the Einstein
field equations, which is a remarkable property of this transformation.

To summarize, the above procedure for obtaining solutions dual to any known 
spacetime solution would work as long as there occurs a free equation in the 
field equations (\ref{vac}) that is not used in finding that solution. Note 
that this holds for all solutions in the family of charged Kerr black 
holes \cite{ND4,DP2} as well as for the NUT solution \cite{NDL}.  
Then the dual  set  admits  a  solution similar to the original one, but with a
topological defect, namely, global monopole charge.

\section{4D dilaton gravity}
\label{sec:dg}

In the spirit of the Barriola-Vilenkin solution (\ref{gms}), one may expect  
analogous solutions to exist even in some scalar-tensor theories of gravity.
A particular class of candidates among these classical theories, which are 
posed as leading alternatives to general relativity, are the 4D 
low-energy effective theories derived from heterotic string theory. Here, we 
consider the specific case of 4D dilaton gravity action coupled 
to a $U(1)$ gauge field, which has charged black hole solutions \cite{GM,GHS}
(see Refs. \cite{rev1,rev2} for reviews):
\be
\label{S4D}
S = {1\over 16\pi} \int d^4 x \>\sqrt{-\bar{g}} \left\{ e^{-2\phi} [\bar{R} +
 4 (\bar{\nabla} \phi)^2 - 2\Lambda ] - \bar{F}^2 \right\} \ \ ,
\ee
where $\bar{g}_{\mu\nu}$ is the string metric, $\bar{R}$ is the 4D Ricci 
scalar, $\Lambda$ is a cosmological constant and $\bar{F}_{\mu\nu}$ is the 
Maxwell field associated with a U(1) subgroup of ${\rm E}_8 \times {\rm E}_8$. 
Here, we shall consider the case where $\Lambda = 0$. The conformal 
transformation $g_{\mu\nu} = e^{-2\phi} \bar{g}_{\mu\nu}$ can be implemented 
to recast the above action in the ``Einstein-Hilbert'' form:
\be
\label{SEH}
S = {1\over 16\pi} \int d^4 x \>\sqrt{-g} \left[ R -2(\nabla 
\phi)^2 - e^{-2\phi}F^2 \right] \ \ ,
\ee
where $g_{\mu\nu}$ is the Einstein-frame metric.
The corresponding equations of motion are:
\begin{mathletters}%
\label{eom}
\begin{eqnarray}
R_{\mu \nu} - {1\over 2}g_{\mu \nu}R -2\nabla_\mu \phi \nabla_\nu \phi + 
g_{\mu \nu} (\nabla \phi)^2 &-& 2e^{-2 \phi} F_{\mu \lambda} F_\nu^\lambda + 
{1\over 2} g_{\mu \nu} e^{-2 \phi} F^2 = 0 \ \ , \label{geom} \\
2\nabla^2 \phi + e^{-2 \phi} F^2 &=& 0 \ \ , \label{peom} \\
\nabla^\nu (e^{-2 \phi}) F_{\mu \nu} &=& 0 \,. \label{Feom}
\end{eqnarray}
\end{mathletters}%
Taking the trace of (\ref{geom}) gives:
\be
\label{Rsca}
R = 2(\nabla \phi)^2 \ \ ,
\ee
which, together with (\ref{geom}) and (\ref{Feom}), implies
\be\label{Rp}
R_{\mu \nu} = 2\nabla_\mu \phi \nabla_\nu \phi + 2e^{-2 \phi}
F_{\mu \lambda} F_\nu^\lambda + g_{\mu \nu} \nabla^2 \phi \,.
\ee
This equation will play a pivotal role below in our study of charged black hole
solutions and their duals. 

If a matter action, say, corresponding to the Lagrangian density (\ref{Lgm}) 
of a global monopole, were present in Eq. (\ref{SEH}), then a matter stress 
tensor following from such a term would contribute to the right-hand side of 
the field equation (\ref{geom}). Since it is not known how the dilaton couples
to the monopole field $\psi^a$, one can obtain different theories incorporating
$\psi^a$ based on the choice of this coupling. Global monopoles in 4D dilaton 
gravity have been studied for some choices of coupling and for both massive 
and massless dilaton in Ref. \cite{DG}. We will later consider solutions to
action (\ref{SEH}) coupled to $\psi^a$ by adding to it the matter action
\be\label{sm}
S_m = -{1\over 16\pi} \int d^4 x \sqrt{-g} \> L_m \ \ ,
\ee
where $L_m$ is given in Eq. (\ref{Lgm}). Such a choice of coupling and $L_m$ is
different from that considered in Ref. \cite{DG}. 

We begin by briefly recalling how the above equations of motion are solved to
obtain the charged black hole solution. Consider the spherically symmetric
static (SSS) metric (\ref{SSS}),
%\be\label{SSS}
%ds^2 = - e^{2\nu} dt^2 + e^{2\lambda} dr^2 + h^2 d \omega^2 \ \ ,\ee
where $\nu$, $\lambda$, and $h$ are now functions of the radial coordinate $r$
only. For such a metric the only non-vanishing components of the Ricci tensor 
are the diagonal elements. For Eq. (\ref{Rp}) to have an SSS solution, the 
following conditions on $R_{tt}$ and $R_{rr}$ must be obeyed:
\begin{mathletters}%
\label{Rtr}
\begin{eqnarray}
R_{tt} =e^{2(\nu-\lambda)}\left\{\nu''+\nu'^2 - \lambda' \nu' +{ 2\nu'\over r}
\right\} -2 \nu' e^{2(\nu-\lambda)} \left[{1\over r} - {h'\over h}\right] &=&
e^{2 \nu - 2 \phi} {Q^2 \over h^4} \ \ , \label{Rtt} \\
R_{rr} =\left\{-\nu-{\nu'}^2 + \lambda'\nu' + {2\lambda'\over r}\right\}
- 2\lambda'\left[{ 1\over r} - {h'\over h}\right] - {2h''\over h} &=& \left[2 
\phi'^2 - e^{2 \lambda -2 \phi} {Q^2\over h^4}\right]\,.\label{Rrr} 
\end{eqnarray}
\end{mathletters}%
Similarly, the field equation for the component $R_{\theta \theta}$,
\be\label{Rtp}
\left\{1+e^{-2 \lambda} \left[r(\lambda'-\nu') -1\right]\right\} +
e^{-2\lambda} \left[(hh'- r) (\lambda'-\nu') - (hh''+h'^2) + 1\right]
= e^{-2 \phi}{Q^2\over h^2} \ \ ,
\ee
must also be satisfied.
Spherical symmetry ensures that the $R_{\phi\phi}$ equation implies the same
condition on $\nu$, $\lambda$, and $h$ as in Eq. (\ref{Rtp}). If $Q=0$, then 
the right-hand side of the above equation vanishes. Therefore, the following 
expressions, 
\be\label{Sch}
e^{2\nu} = e^{-2 \lambda} = 1-2m/r \>, \quad {\rm and} \quad h = r 
\ee
constitute a solution to the above equations. This simply corresponds to the 
Schwarzschild metric, which indeed is a solution to the equations of motion
(\ref{eom}) with $F_{\mu\nu} = 0$ and $\phi =$ constant. 

Finding an SSS metric 
as a solution to (\ref{Rtr}) and (\ref{Rtp}) for $Q\neq 0$ is also 
straightforward. Note that $\nu$ and $\lambda$ given in Eq. (\ref{Sch}) makes
the braces on the left-hand sides of Eqs. (\ref{Rtr}) and (\ref{Rtp}) vanish. 
Thus, the problem reduces to finding an $h$ that makes the 
remaining term on the left-hand sides of Eqs. (\ref{Rtr}) and (\ref{Rtp}) equal
to their right-hand sides, respectively. Such an $h$ exists and is given by 
\be\label{h}
h^2 = r^2\left( 1-{Q^2\over mr}\right) \,.
\ee
This, therefore, constitutes the charged black hole solution of 4D dilatonic
gravity. The corresponding fields are:
\begin{eqnarray}
{ds}^2 &=& -\left( 1-{2m \over r}\right){dt}^2 +
\left(1- {2m \over r}\right)^{-1} {dr}^2 \no \\
&&+ r^2 \left( 1- {Q^2 \over mr}\right){d\omega}^2 \,. \label{GHSE}\\
e^{-2\phi} &=& \left( 1 - {Q^2 \over mr}
 \right) = U(\phi)\ \ ,\label{4Ddil} \\
F_{rt}&=&{Q\over r^2} \,. \label{4DF}
\end{eqnarray}
In obtaining the above solution from the equations of motion, we have assumed
that $\phi\to 0$ as $r\to \infty$.

The effect of the electrogravity-duality transformation on the field equations 
(\ref{geom}) is to modify them by the addition of the asymptotic form of the
global-monopole stress 
tensor (\ref{gmstress}) on its right-hand side. This is completely analogous to
what happens in the case of the Schwarzschild black hole in GR (see section
\ref{sec:sd}).
It, therefore, follows that the following new choices for $\nu$ and 
$\lambda$ will solve such a set of equations:
\begin{eqnarray} \label{nmg}
e^{2\nu}\to e^{2\tilde{\nu}} &=&  1-8\pi\eta^2 -{2\tilde{m}\over r} \ \ , \no\\
e^{2\lambda}\to e^{2\tilde{\lambda}}&=& 1-8\pi\eta^2 -{2\tilde{m}\over r} \ \ ,
\end{eqnarray}
where $\tilde{m}$ is just an integration constant. In other words, for the 
choice in (\ref{nmg}),
the braces on the lhs of Eqs. (\ref{Rtr}) and (\ref{Rtp}) are exactly equal to 
the components of the global-monopole stress-tensor. This suggests the 
possibility that there exists an $h\to \tilde{h}$ and $\phi \to\tilde{\phi}$, 
for which $\tilde{\nu}$ and $\tilde{\lambda}$ solve the field equations 
(\ref{Rtr}) and  (\ref{Rtp}) modified by the presence of source terms arising
from the global monopole stress tensor.

In fact, it turns out that such a choice for $h$ is available. This 
can be understood by noting that $\tilde{\nu}$ and $\tilde{\lambda}$ can be 
cast in the same form as (\ref{Sch}):
\be \label{newoldnmg}
e^{2\tilde{\nu}}= e^{-2\kappa} \left(1- {2m\over r}\right) = e^{-2
\tilde{\lambda}} \ \ ,
\ee
where $e^{-2\kappa} = (1-8\pi\eta^2)$ and $\tilde{m}= e^{-2\kappa}m$. Using 
such scaling relations between tilded and untilded parameters, it is easy to 
see that 
\be\label{tilhp2}
\tilde{h}^2 = r^2 \left(1-{Q^2\over \tilde{m}r}\right)\>,\quad {\rm and} \quad
e^{-2\tilde{\phi}}= \left(1-{Q^2\over \tilde{m}r}\right) \,.
\ee
Calling $\tilde{m} = M$, we finally arrive at the metric of the spacetime 
dual to the charged dilatonic black holes:
\be\label{gmmet}
ds^2 = - \left(1-8 \pi\eta^2 -{2M\over r}\right) dt^2 +
\left(1- 8\pi \eta^2 - {2M\over r}\right)^{-1} dr^2 +
r^2 \left(1- {Q^2\over Mr}\right){d\omega}^2 \,.
\ee
This is a solution to the modified equations of motion. The corresponding 
dilaton and $U(1)$ field solutions are given by Eqs. (\ref{tilhp2}) and
(\ref{4DF}), respectively. The resulting field configuration solves the 
equations of motion (\ref{eom}). In the limit $Q= 0$, one recovers the 
Barriola-Vilenkin spacetime. This is expected since in that limit the dilaton 
field acquires a constant value. Consequently, 4D dilaton gravity reduces to
general relativity. Additionally, if $M=0$, then the above metric describes a
locally flat spacetime with a global monopole charge, which is the 
electrogravity dual of flat spacetime.

It, however, remains to be shown that the stress tensor (\ref{gmstress}) is
indeed the asymptotic form of the global monopole stress tensor arising from 
(\ref{sm}) for the above spacetime metric.  
To see that this is indeed true, we cast the above metric as 
\begin{eqnarray}
ds^2 = &-&\left( 1-{4M^2\over Q^2 +\sqrt{1+{4M^2 \rho^2
\over Q^4}}} \right) dt^2 \no \\
&+&{1\over 4} \left(1+ {Q^4\over 4M^2 \rho^2 } \right)^{-1} 
\left( 1-{4M^2\over Q^2 + \sqrt{1+{4M^2 \rho^2 \over 
Q^4}}} \right)^{-1} d\rho^2 +\rho^2 d\omega^2 \ \ ,\label{arR}
\end{eqnarray}
where $\rho^2= r^2 - rQ^2/ M$. This metric is of the same form as Eq. 
(\ref{SSS}) with $r$ replaced by $\rho$ there. Using Eqs. 
(\ref{gmst}) to compute the matter stress-tensor components gives 
\be\label{gmstd}
T_t{}^t\approx T_\rho{}^\rho \approx \eta^2/\rho^2 \>, \quad T_\t{}^\t = 
T_\phi{}^\phi \approx 0
\ \ , \ee
outside the monopole core. In the limit of large $r$ these go over to the 
expected stress tensor components (\ref{emt}) of a global monopole. Moreover,
it is straightforward to verify that the field equation for $f(\rho)$, which is
given by Eq. (\ref{fr}) with $r$ replaced by $\rho$, can be solved 
asymptotically with $f(\rho)\approx 1$ outside the core.

\section{Brans-Dicke theory}
\label{sec:bd}

Another alternative candidate for the theory of gravity is the Brans-Dicke 
theory (BD). Analogous to the dilaton field in string theory, BD also includes
a scalar field as part of the spacetime geometry. The BD action is, however,
different from those considered in the previous sections \cite{MTW}. Vacuum 
BD can be conformally transformed into 4D Einstein gravity coupled to a 
massless scalar field, $\varphi$. The corresponding field equations are:
\be\label{BDfe}
R_{\mu\nu} -{1\over 2} Rg_{\mu\nu} = \kappa (T^\varphi_{\mu\nu} +T_{\mu\nu}) 
\ \ , \ee
where $\kappa$ is a coupling constant, $T^\varphi_{\mu\nu}$ is the stress 
tensor of the scalar field,
\be\label{BDTs}
T^\varphi_{\mu\nu} = \nabla_\m \varphi\nabla_\n \varphi -{1\over 2}g_{\mu\nu}
\nabla^\lambda \varphi\nabla_\lambda \varphi
\ \ , \ee
and $T_{\mu\nu}$ is the stress tensor of any matter distribution that may be 
present. Also, the scalar field obeys the equation,
\be\label{BDsfe}
\nabla^\m \nabla_\m \varphi = 0 \ \ ,
\ee
Equations (\ref{BDfe}) and (\ref{BDTs}) can be combined to yield
\be\label{BDR}
R_{\m\n} = \kappa (\nabla_\m \varphi\nabla_\n \varphi +T_{\mu\nu})\,.
\ee
Thus, Eqs. (\ref{BDR}) and (\ref{BDsfe}) comprise the field equations of this
theory. Note that these equations can be obtained from the more general class
of field equations, Eqs. (\ref{eom}) by setting $F_{\m\n}=0$, $\kappa = 2$, and
identifying the dilaton field with the scalar field in those equations. 

The spherically symmetric neutral black hole solutions to the above equations 
are known. The no scalar-hair theorem \cite{Bek1,Saa} guarantees that they are 
just the Schwarzschild solution with a constant scalar field \cite{Hawk1,XZ}. 
Also, the spherically symmetric charged black hole solution to the 4D 
Brans-Dicke-Maxwell theory still remains the Reissner-Nordstrom solution with 
a constant scalar field \cite{Cai}. However, as was shown by Xanthopoulos and 
Zannias (XZ) \cite{XZ}, there is an interesting solution to the above 
equations with a varying scalar field and with a vanishing matter stress 
tensor, $T_{\mu\nu}$. The corresponding fields are:
\be\label{XZst}
ds^2 = -dt^2 + dr^2 + (r^2 -r_0^2) d\omega^2 \ \ ,
\ee
and
\be\label{XZs}
\varphi(r) = {1\over\sqrt{2\kappa}} \ln\left[r-r_0\over r+r_0\right] \ \ ,
\ee
where $r_0$ is a constant. The scalar curvature behaves as:
\be
R= {2r_0\over (r^2-r_0^2)} \,.
\ee
Although not a black hole, nevertheless the above solution has a naked 
curvature singularity at $r=r_0$. We call this the XZ solution. Below, 
we seek the dual of this solution.

Under EGT, the field equation (\ref{BDfe}) gets transformed to:
\be\label{dBDfe}
G_{\mu\nu} -{1\over 2} Rg_{\mu\nu} = \kappa (T^\varphi_{\mu\nu} +T_{\mu\nu}) 
\,. \ee
It can be shown that the spacetime solution to the above equations is:
\be\label{XZd}
ds^2 = -(1-8\pi\eta^2) dt^2 + (1-8\pi\eta^2)^{-1}dr^2 +(r^2 -r_0^2) 
d\omega^2 \,. \ee
The above form of the dual metric is expected since the original 
metric is spherically symmetric. The only non-vanishing component of the 
Einstein tensor for the above metric is
\be\label{BDdGrr}
G_r{}^r = {2r_0^2\over (1-8\pi\eta^2)(r^2-r_0^2)} \,.
\ee 
The corresponding scalar field solution is the same as that given by 
(\ref{XZs}). This is because, under the duality transformation, the 
determinant of the XZ metric (\ref{XZst}) remains invariant and $g^{rr}$ 
continues to remain constant. 

The matter stress tensor, $T_{\mu\nu}$, for this solution is non-vanishing; 
in fact, it is the same as that for a global monopole (\ref{emt}), as is 
typical of solutions obtained by applying EGT on known SSS solutions. Indeed, 
in the limit $r_0  \to 0$, the metric (\ref{XZd}) clearly describes a space 
with a deficit solid angle. This limit corresponds to a constant scalar field 
and, therefore, yields a solution in GR. Such a solution was indeed observed 
in Ref. \cite{BV}, and is associated with a global monopole.

\section{Discussion}
\label{sec:disc}

An important observation made in Refs. \cite{ND2,ND3} was that as long as one 
of the Einstein field equations remains unused in obtaining a particular
solution, one can obtain a dual solution, which is different from the original,
by modifying that equation. The modification is to introduce the term on the 
right in Eqs. (\ref{modvac}), where the spacelike unit vector $w_a$ is along
the radial four-acceleration vector. By this prescription solutions dual to 
all isolated sources have been obtained. The next question is: How good are 
Eqs. (\ref{modvac}) as ``non-vacuum'' field equations? The first part of the 
equation implies vanishing of energy density (i.e., $\tilde E = 0$) and of 
energy flux (i.e., $H_{[ab]} = 0$). This means that they cannot have a 
physically meaningful non-vacuum solution. In the case of spacetimes 
corresponding to non-localized sources, such as those associated with the 
Kasner, the Weyl, the Levi-Civita metrics, and the metric of a plane 
gravitational wave,  it turns out that Eqs. (\ref{modvac}) (with $k w_a 
w_b$ replaced by $k (g_{ab} + u_au_b)$ for the homogeneous case) admits 
them as solutions, and so does the dual set of equations \cite{PD}. Thus, such
solutions describing non-localized sources are electrogravity self-dual. Also,
Eqs. (\ref{modvac}) admit meaningful solutions only when they are vacuum 
spacetimes. Even if they admit any non-vacuum solution, it would correspond to
a matter distribution with vanishing energy density and, hence, would be
unphysical. Hence, although the electrogravity duality transformation can be
effected to obtain dual spacetime solutions in the above manner, it is not 
guaranteed to correspond to physically acceptable matter distributions. For
instance, there does not occur a  physically reasonable dual solution to any 
spherically symmetric static perfect fluid spacetime other than the de Sitter 
solution. 

The application of the electrogravity-dual transformation on perfect fluid 
spacetimes has been considered \cite{DPT}. It maps the perfect-fluid 
stress-tensor components as $\rho\to (\rho+3p)/2$ and $p\to (\rho-p)/2$, where
$\rho$ and $p$ are the matter energy and pressure densities, respectively. 
Thus, the transformed stress tensor also describes a perfect fluid. In 
particular, the equations of state corresponding to radiation, $\rho=3p$, and 
to the cosmological constant term, $\rho+p=0$, remain invariant. In the former 
case, the solution is self-dual because the Ricci tensor is identical to the 
Einstein tensor. Whereas in the latter case we have $\Lambda \to -\Lambda$, 
which implies that the de Sitter and the anti-de Sitter spacetimes are dual to 
each other. Also, the dust distribution, where $p=0$, is dual to the stiff
fluid, where $\rho =p$.

Since electrogravity duality transformation only involves the Riemann tensor 
and hence is quite general, it should be applicable in other metric theories 
as well. Here we have seen its application on black hole solutions in the 4D 
low-energy effective heterotic string theory as well as on the 
Xanthopoulos-Zannias solution in Brans-Dicke theory. The procedure works 
exactly along the lines of GR and we obtain their dual solutions, which 
exhibit the presence of a global monopole charge. Thus we can make the general 
statement that {\it by implementing the electrogravity duality transformation 
one can always generate a global monopole charge in a static spherically 
symmetric solution in GR, Brans-Dicke theory, and in 4D dilaton gravity}. This
is because as in GR, even in these scalar-tensor theories, the static 
spherically symmetric sector of their solution space is determined
by only a subset of the corresponding field equations. In fact this is true
even in lower dimensional Einstein-gravity, e.g., the theory corresponding to
the 3D Einstein-Hilbert action involving a cosmological constant
term. Hence, these theories, which have played an important role in this decade
in the understanding of black hole physics and related quantum aspects, also
hold the promise of harboring yet unknown solutions with topological defects
that may play an important role in alternative cosmological models. We are
currently involved in studying such solutions \cite{BDK}.

\section{Acknowledgments}

We thank Sayan Kar for his critical reading of our notes and for helpful 
discussions.

\end{document}